\documentclass[12pt]{extarticle}
\usepackage{setspace}

\usepackage{setspace}
\usepackage[T1]{fontenc}
\usepackage{times}
\usepackage{palatino}
\usepackage{lmodern}
\usepackage[latin1]{inputenc}
\usepackage{epsfig}
\usepackage[english]{babel}
\usepackage{color}
\usepackage{graphicx}
\usepackage{dcolumn}
\usepackage{moreverb}
\usepackage{amsmath,amssymb,amsfonts}
\usepackage[all]{xy}

\begin{document}
\numberwithin{equation}{section}
\newcommand{\boxedeqn}[1]{%
  \[\fbox{%
      \addtolength{\linewidth}{-2\fboxsep}%
      \addtolength{\linewidth}{-2\fboxrule}%
      \begin{minipage}{\linewidth}%
      \begin{equation}#1\end{equation}%
      \end{minipage}%
    }\]%
}


\newsavebox{\fmbox}
\newenvironment{fmpage}[1]
     {\begin{lrbox}{\fmbox}\begin{minipage}{#1}}
     {\end{minipage}\end{lrbox}\fbox{\usebox{\fmbox}}}

\raggedbottom
\onecolumn

\parindent 8pt
\parskip 10pt
\baselineskip 16pt
\noindent\title*{{\Large{\textbf{Quadratic algebra structure in the 5D Kepler system with non-central potentials and Yang-Coulomb monopole interaction}}}}
\newline
\newline
Md Fazlul Hoque $^a$, Ian Marquette $^a$ and Yao-Zhong Zhang $^{a,b}$
\newline
\newline
$^a$ School of Mathematics and Physics, The University of Queensland, Brisbane, QLD 4072, Australia
\newline
\newline
$^b$ CAS Key Laboratory of Theoretical Physics, Institute of Theoretical Physics, Chinese Academy of Sciences, Beijing 100190, China
\newline
\newline
{\tt  m.hoque@uq.edu.au, i.marquette@uq.edu.au, yzz@maths.uq.edu.au}
\newline
\newline

\begin{abstract}
 We construct the integrals of motion for the 5D deformed Kepler system with non-central potentials in $su(2)$ Yang-Coulomb monopole field. We show that these integrals form a higher rank quadratic algebra $Q(3; L^{so(4)}, T^{su(2)})\oplus so(4)$, with structure constants involving the Casimir operators of $so(4)$ and $su(2)$ Lie algebras. We realize the quadratic algebra in terms of the deformed oscillator and construct its finite-dimensional unitary representations. This enable us to derive the energy spectrum of the system algebraically. Furthermore we show that the model is multiseparable and allows separation of variables in the hyperspherical and parabolic coordinates. We also show the separability of its 8D dual system (i.e. the 8D singular harmonic oscillator) in the Euler and cylindrical coordinates.
\end{abstract}

\section{Introduction}

 Quantum mechanical systems in Dirac monopole \cite{Dir1} or Yang's non-abelian $su(2)$ monopole \cite{Yan1} fields have remained a subject of international research interest. In \cite{Tru1} the structural similarity between  $su(2)$ gauge system and the 5D Kepler problem was discussed in the scheme of algebraic constraint quantization. Various 5D Kepler systems with $su(2)$ monopole (or Yang-Coulomb monopole) interactions have been investigated by many authors \cite{Sis1, Mad1, Mad2, Ple3, Ple1, Mad5, Ple2, Kal1, Ner1, Mad12, Kar1, Bel1, Bel2}. Superintegrability  of certain monopole systems has been shown in \cite{Mad3, Mad4, Ran1, Sal1, Mar1, FH1, FH2}.

There exists established connection between  2, 3, 5 and 9-dimensional hydrogen-like atoms and 2, 4, 8 and 16-dimensional harmonic oscillators, respectively
\cite{Lev1, Dav1, Kle1, Le1, Van1, Van2}.  In \cite{Ian1} one of the present authors proved the superintegrability of the 5D Kepler system deformed with non-central terms in Yang-Coulomb monopole (YCM) field. This was achieved by relating this system to an 8D singular harmonic oscillator via the Hurwitz transformation \cite{Hur1}. An algebraic calculation of the spectrum of the 8D dual system was performed, which enabled the author to deduce the spectrum of the 5D deformed Kepler YCM system via duality between the two models.

However, algebraic structure of the 5D deformed Kepler YCM system and a direct derivation of its spectrum via the symmetry algebra have remained an open problem. Such algebraic approach is expected to provide deeper understanding to the degeneracies of the energy spectrum and the connection of the wavefunctions with special functions and orthogonal polynomials.

Symmetry algebras generated by integrals of motion of superintegrable systems are in general polynomial algebras with structure constants involving Casimir operators of certain Lie algebras \cite{Gra1, Das1, Kal2, Kal3,  Das3, Mil1, Gen1, Isa1, FH3, FH4, FH5}.
The purpose of this paper is to obtain the algebraic structure in the 5D deformed Kepler YCM system  and apply it to give a direct algebraic derivation of the energy spectrum of the model. We construct the integrals of motion and show that they form a higher rank quadratic algebra $Q(3; L^{so(4)}, T^{su(2)})\oplus so(4)$ with structure constants involving the Casimir operators of $so(4)$ and $su(2)$ Lie algebras. This quadratic algebra structure enable us to give an algebraic derivation of the energy spectrum of the model. Furthermore, we show that the model is separable in the hyperspherical and parabolic coordinates. We also show the separability of its 8D dual system (i.e. the 8D singular harmonic oscillator) in Euler and cylindrical coordinates.

\section{Kepler system in Yang-Coulomb monopole field}

The 5D Kepler-Coulomb system with Yang-Coulomb monopole interaction is defined by the Hamiltonian \cite{Sis1, Ian1}
\begin{eqnarray}
H=\frac{1}{2}\left(-i\hbar \frac{\partial}{\partial x_{j}} -\hbar A_{j}^{a} T_{a}\right)^{2}  +\frac{\hbar^{2}}{2r^{2}} \hat{T}^2 -\frac{c_{0}}{r},\label{H-undeformed}
\end{eqnarray}
where $j=0, 1, 2, 3, 4$ and $\{T_a, a=1, 2, 3\}$ are the $su(2)$ gauge group generators which satisfy the  commutation relations
\begin{eqnarray}
[T_a, T_b]=i\epsilon_{abc}T_c
\end{eqnarray}
$\hat{T}^2=T^a T^a$ is the Casimir operator of $su(2)$ and $A^a_j$ are the components of the monopole potential $\textbf{A}^a$, which can be written as
\begin{eqnarray}
A_j^a=\frac{2i}{r(r+x_0)}\tau^a_{jk}x_k.
\end{eqnarray}
Here $\tau^a$ are the $5\times 5$ matrices
\begin{eqnarray}
\tau^1 =\frac{1}{2}
\begin{pmatrix}
0 & 0 & 0 \\
0 & 0 & -i\sigma^1 \\
0 & i\sigma^1 & 0
\end{pmatrix},\quad
\tau^2 =\frac{1}{2}
\begin{pmatrix}
0 & 0 & 0 \\
0 & 0 & i\sigma^3 \\
0 & -i\sigma^3 & 0
\end{pmatrix},\quad
\tau^3 =\frac{1}{2}
\begin{pmatrix}
0 & 0 & 0 \\
0 & \sigma^2 & 0 \\
0 & 0 &  \sigma^2
\end{pmatrix},\nonumber\\
\end{eqnarray}
which satisfy $[\tau^a, \tau^b]=i\epsilon_{abc}\tau^c$ and $\sigma^i$ are the Pauli matrices
\begin{eqnarray}
\sigma^1 =
\begin{pmatrix}
0 & 1 \\
1 & 0
\end{pmatrix},\quad
\sigma^2 =
\begin{pmatrix}
0 & -i \\
i & 0
\end{pmatrix},\quad
\sigma^3 =
\begin{pmatrix}
1 & 0 \\
0 & -1
\end{pmatrix}.
\end{eqnarray}
The vector potentials $\textbf{A}^a$ are orthogonal to each other,
\begin{eqnarray}
\textbf{A}^a\cdot \textbf{A}^b=\frac{1}{r^2}\frac{r-x_0}{r+x_0}\delta_{ab}
\end{eqnarray}
and to the vector $\textbf{x}=(x_0, x_1, x_2, x_3, x_4)$.

The Kepler-Coulomb model Hamiltonian has the following integrals of motion \cite{Yan1}
\begin{eqnarray}
&& L_{jk}=(x_j\pi_k-x_k\pi_j)-r^2\hbar F^a_{jk}T_a,\nonumber\\
&& M_{k}=\frac{1}{2}(\pi_jL_{jk}+L_{jk}\pi_j)+c_0\frac{x_k}{r},\label{ar1}
\end{eqnarray}
which satisfy $[H, L_{jk}]=0=[H, M_{k}]$. Here
\begin{eqnarray}
F^a_{jk}=\partial_j A^a_k-\partial_k A^a_j+\epsilon_{abc}A^b_j A^c_k.\label{f1}
\end{eqnarray}
is the Yang-Mills field tensor and $\pi_j=-i\hbar \frac{\partial}{\partial x_{j}} -\hbar A_{j}^{a} T_{a}$ which obey the commutation relations
\begin{eqnarray}
[\pi_j, x_k]=-i\hbar\delta_{jk}, \quad [\pi_j,\pi_k]=i\hbar^2 F^a_{jk} T_a.
\end{eqnarray}
The integrals $L_{jk}$ and $M_k$ close to the  $so(6)$ algebra  \cite{Lan1}
\begin{eqnarray}
&&[L_{ij}, L_{mn}]=i\hbar\delta_{im}L_{jn}-i\hbar\delta_{jm}L_{in}-i\hbar\delta_{in}L_{jm}+i\hbar\delta_{jn}L_{im},\nonumber\\
&&[L_{ij}, M_{k}]=i\hbar\delta_{ik}M_{j}-i\hbar\delta_{jk}M_{i}, \quad [M_{i}, M_{k}]=-2i\hbar H L_{ik}.
\end{eqnarray}
The $so(6)$  symmetry algebra is very useful in deriving the energy spectrum of the system algebraically.

The Casimir operators of $so(6)$ are given by \cite{Bar1, Mad11}
\begin{eqnarray}
&&\hat{K}_1=\frac{1}{2}D_{\mu\nu}D_{\mu\nu}, \quad\quad \hat{K}_2=\epsilon_{\mu\nu\rho\sigma\tau\lambda} D_{\mu\nu}D_{\rho\sigma}D_{\tau\lambda},\nonumber \\
&& \hat{K}_3=\frac{1}{2}D_{\mu\nu}D_{\nu\rho}D_{\rho\tau}D_{\tau\mu},
\end{eqnarray}
where $\mu, \nu= 0,1,2,3,4$ and
\begin{eqnarray}
D =
\begin{pmatrix}
L_{jk} & -(-2H)^{1/2}M_k \\
(-2H)^{1/2}M_k & 0
\end{pmatrix}.
\end{eqnarray}
The eigenvalues of the operators $\hat{K}_1, \hat{K}_2$ and $\hat{K}_3$ are  \cite{Per1}
\begin{eqnarray}
&&K_1=\mu_1(\mu_1+4)+\mu_2(\mu_2+2)+\mu_3^2,\nonumber\\
&&K_2=48(\mu_1+2)(\mu_2+1)\mu_3,\nonumber \\
&& K_3=\mu_1^2(\mu_1+4)^2+6\mu_1(\mu_1+4)+\mu_2^2(\mu_2+2)^2+\mu_3^4-2\mu_3^2,
\end{eqnarray}
respectively, where $\mu_1$, $\mu_2$ and $\mu_3$ are positive integers or half-integers and $\mu_1\geq\mu_2\geq\mu_3$. One can represent the operators $\hat{K}_1, \hat{K}_2, \hat{K}_3$ in the form \cite{Mad11}
\begin{eqnarray}
&&\hat{K}_1=-\frac{c_0}{2\hbar^2 H}+2\hat{T}^2-4, \quad\quad \hat{K}_2=48\left(-\frac{\mu_0}{2\hbar^2 H}\right)^{1/2}\hat{T}^2,\nonumber \\
&& \hat{K}_3=K_1^2+6K_1-4K_1 \hat{T}^2-12\hat{T}^2+6\hat{T}^4.
\end{eqnarray}
Denote by $T$ the eigenvalue of $\hat{T}^2$. Then,
\begin{eqnarray}
&&K_1-2T(T+1)=\mu_1(\mu_1+4),\nonumber  \\
&&\mu_2^2(\mu_2+2)^2+\mu_3^4-2\mu_3^2=2T^2(T+1)^2,
\end{eqnarray}
The energy levels of the Yang-Coulomb monopole system are given by
\begin{eqnarray}
\epsilon^{T}_n=-\frac{c_0}{2\hbar^2(\frac{n}{2}+2)^2},
\end{eqnarray}
where use has been made of $\mu_1=\frac{n}{2}$ with $n$ being nonnegative integer.

\section{Kepler system with non-central potentials and Yang-Coulomb monopole interaction}
Let us now consider the 5D Kepler system deformed with non-central potentials in Yang-Coulomb monopole field. The Hamiltonian is given by \cite{Yan1, Sis1, Ian1}
\begin{eqnarray}
&&H=\frac{1}{2}\left(-i\hbar \frac{\partial}{\partial x_{j}} -\hbar A_{j}^{a} T_{a}\right)^{2}  +\frac{\hbar^{2}}{2 r^{2}} \hat{T}^2 -\frac{c_{0}}{r} + \frac{c_{1}}{r(r+x_{0})} + \frac{c_{2}}{r(r-x_{0})},\nonumber\\&&\label{KP1}
\end{eqnarray}
where $c_1$ and $c_2$ are positive real constants. We can construct the integrals of motion of (\ref{KP1}) by deforming the integrals (\ref{ar1}) of (\ref{H-undeformed}),
\begin{eqnarray}
&&A= \hat{L}^2+ \frac{2 r c_{1}}{(r+x_{0})} + \frac{2 r c_{2}}{(r-x_{0})},\nonumber\\
&&B=M_k+ \frac{c_{1}(r-x_{0})}{\hbar^{2} r(r+x_{0})}  + \frac{c_{2}(r+x_{0})}{\hbar^{2} r(r-x_{0})},\label{IntegralsAB}
\end{eqnarray}
where
\begin{eqnarray}
\hat{L}^2=\sum_{i<j}{L}^2_{ij}, \quad\quad L_{ij}=x_i p_j-x_j p_i,\quad i,j=1, 2, 3, 4.
\end{eqnarray}
It can be checked that these integrals satisfy the commutation relations
\begin{eqnarray}
&[H,A]=0=[H,B],\quad\quad [A,\hat{L}^2]=0=[B,\hat{L}^2],& \nonumber
\\& [H,\hat{L}^2]=0=[H,L_{jk}].&
\end{eqnarray}
So the 5D deformed Kepler YCM system (\ref{KP1}) is minimally superintegrable as it has 6 algebraically independent integrals of motion including $H$.

The integrals $A, B$ (\ref{IntegralsAB}) have the following differential operator realization:
\begin{eqnarray}
&A&=\hbar^2 \left[-r^{2}\frac{\partial^2}{\partial x_j^2} +x_{j}x_{k} \frac{\partial^{2}}{\partial x_{j} \partial x_{k}}+4 x_{j} \frac{\partial}{\partial x_{j}} \frac{2 r}{r+x_{0}} \hat{T}^2  \right.\nonumber\\
&&\quad\left. + 2 i r^{2} A_{j}^{a} T_{a} \frac{\partial}{\partial x_{j}}+ \frac{2 r c_{1}}{(r+x_{0})} + \frac{2 r c_{2}}{(r-x_{0})}\right],\nonumber\\
&B&=\hbar^2 \left[ x_{0} \frac{\partial^{2}}{\partial x_j^2 }- x_j \frac{ \partial^{2}}{\partial x_{0} \partial x_j}+i(r-x_{0})A_{j}^{a} T_{a} \frac{\partial}{\partial x_{j}} -2 \frac{\partial}{\partial x_{0}} \right.\nonumber\\
&&\quad\left.+ \frac{ (r-x_{0})}{r(r+x_{0})}\hat{T}^2 +\frac{c_{0}}{\hbar^{2}}\frac{x_{0}}{r}\right]+ \frac{c_{1}(r-x_{0})}{\hbar^{2} r(r+x_{0})}  + \frac{c_{2}(r+x_{0})}{\hbar^{2} r(r-x_{0})}.\label{DifferentialAB}
\end{eqnarray}
These differential expressions are associated with the multiseparability of the Hamiltonian, as will be seen later. However, let us point out that in general systems with monopole interactions are not separable \cite{Win1}.

\section{Algebra structure, unirreps and energy spectrum}
By direct computation, we can show that the integrals $A$, $B$ (\ref{IntegralsAB}) and the central elements $H$, $\hat{L}^2$, $\hat{T}^2$ close to form the following quadratic algebra
$Q(3; L^{so(4)}, T^{su(2)})$,
\begin{eqnarray}
&&[A,B]=C,\label{q1}
\\&&[A,C]=2 \{A,B\} +8 B -2c_{0}(c_{1}-c_{2})-4 c_{0} \hat{T}^2,\label{q2}
\\&&[B,C]=-2 B^{2} +8 H A - 4 \hat{L}^2 H + (16 -4c_{1}-4c_{2})H +2 c_{0}^{2}.\label{q3}
\end{eqnarray}
The Casimir operator is a cubic combination of the generators, given explicitly by
\begin{eqnarray}
\hat{K}&=&C^2-2\{A,B^2\}-4B^2-2\{4c_0(c_2-c_1)-4c_0 \hat{T}^2\}B+8HA^2 \nonumber\\&&+2\{(16-8c_1-8c_2)H-4\hat{L}^2H+2c_0^2\}A. \label{C1}
\end{eqnarray}
Using the differential realization (\ref{DifferentialAB}) of the integrals $A$ and $B$,  we can show that the Casimir operator (\ref{C1}) takes the form,
\begin{eqnarray}
\hat{K}&=&-8 H (\hat{T}^2)^{2} +16 \hat{L}^2 H -8(c_{1}-c_{2}) \hat{T}^2 H-2\{(c_{1}-c_{2})^{2} \nonumber\\
&&+8 (2-c_{1}-c_{2})\}H + 4 c_{0}^{2}\hat{L}^2 +4c_{0}^{2}(c_{1} + c_{2}-1).
\end{eqnarray}
Notice that the first order integrals of motion $\{L_{ij},~ i, j=1,2,3,4\}$ generate the $so(4)$ algebra
\begin{eqnarray}
[L_{ij}, L_{mn}]=i\hbar(\delta_{im}L_{jn}-\delta_{jm}L_{in}-\delta_{in}L_{jm}+\delta_{jn}L_{im}).
\end{eqnarray}
So the full dynamical symmetry algebra of the Hamiltonian (\ref{KP1}) is a direct sum $Q(3; L^{so(4)}, T^{su(2)})\oplus so(4)$ of the quadratic algebra $Q(3; L^{so(4)}, T^{su(2)})$ and the $so(4)$ Lie algebra, with structure constants involving the Casimir operators $\hat{L}^2, \hat{T}^2$ of $so(4), su(2)$.

In order to obtain the energy spectrum of the system, we now construct a realization of the quadratic algebra $Q(3; L^{so(4)}, T^{su(2)})$ in terms of the deformed oscillator algebra of the form
\cite{Das1, Das2},
\begin{eqnarray}
[\aleph,b^{\dagger}]=b^{\dagger},\quad [\aleph,b]=-b,\quad bb^{\dagger}=\Phi (\aleph+1),\quad b^{\dagger} b=\Phi(\aleph).\label{kpfh}
\end{eqnarray}
Here $\aleph $ is the number operator and $\Phi(x)$ is well behaved real function satisfying
\begin{eqnarray}
\Phi(0)=0, \quad \Phi(x)>0, \quad \forall x>0.\label{kpbc}
\end{eqnarray}
It is non-trivial to obtain such a realization and to find the structure function $\Phi(x)$. After long computations, we get
\begin{eqnarray}
A&=&(\aleph+u)^2-\frac{9}{4},\nonumber\\
B&=&b(\aleph)+b^{\dagger}\rho(\aleph)+\rho(\aleph)b,
\end{eqnarray}
where
\begin{eqnarray}
b(\aleph)&=&\frac{c_0(c_1-c_2)+c_0 \hat{T}^2}{(\aleph+u)^2-\frac{1}{4}},\nonumber\\
\rho(\aleph)&=&\frac{1}{3. 2^{20}(\aleph+u)(1+\aleph+u)\{1+2(\aleph+u)^2\}},
\end{eqnarray}
and $u$ is a constant to be determined from the constraints on the structure function $\Phi$. Using (\ref{q1}-\ref{q3}) and  (\ref{C1}), we find
\begin{eqnarray}
&\Phi(x;u,H)&=98304 [2 c_0^2 + H \{1 - 2 (x+u)\}^2] [4 c_1^2 +
   4 c_2^2 + \{1 - 2 (x+u)\}^2 \nonumber \\&&\times\{4(x+u)(x+u-1)-4\hat{L}^2-3\} -
   4 c_1 [2 c_2 + \{1 - 2 (x+u)\}^2 \nonumber \\&& - 4\hat{T}^2] + 16 (\hat{T}^2)^2 -
   4 c_2 [\{1 - 2 (x+u)\}^2 + 4 \hat{T}^2]].
   \end{eqnarray}
A set of appropriate quantum numbers can be defined in same way as in \cite{Tru1, Ras1}. We can use the subalgebra chains $so(4)\supset so(3)\supset so(2)$ and $so(3)\supset so(2)$ for $\hat{L}^2$ and $\hat{T}^2$ respectively. The Casimir operators $\hat{J}^2_{(\alpha)}$ of the related chains can be written as \cite{Ras1}
\begin{eqnarray}
\hat{J}_{(\alpha)}^2= \sum^\alpha_{i<j} J^2_{ij}, \quad \alpha = 2, 3, 4.
\end{eqnarray}
We need to use an appropriate Fock space in order to obtain finite-dimensional unitary irreducible representations (unirreps) of $Q(3; L^{so(4)}, T^{su(2)})$. Let $|n,E>\equiv |n, E, l_4, l_3, l_2>$ donote the Fock basis states, where $l_4, l_3, l_2$ are quantum numbers of $\hat{J}_{(\alpha)}^2,~ \alpha=4, 3, 2$, respectively.
Then the eigenvalues of the Casimir operators $\hat{L}^2$ and $\hat{T}^2$ are $\hbar^2 l_4(l_4+2)$ and $\hbar^2 T(T+1)$, respectively.  By acting the structure function on the Fock states $|n, E\rangle$ with $\aleph|n, E\rangle =n|n,E\rangle$ and using the eigenvalues of $H$, $\hat{L}^2$ and $\hat{T}^2$, we get
\begin{eqnarray}
&\Phi(x;u,E)&=[ x+u -(\frac{1}{2}-\frac{c_{0}}{\sqrt{-2E}})][ x+u-(\frac{1}{2}+\frac{c_{0}}{\sqrt{-2E}})]\nonumber\\&& \times  [x+u-\frac{1}{2}(1+m_{1}+m_{2})][x+u-\frac{1}{2}(1+m_{1}-m_{2})]\nonumber\\&& \times [x+u-\frac{1}{2}(1-m_{1}+m_{2})][x+u-\frac{1}{2}(1-m_{1}-m_{2})],\nonumber\\&&\label{pp1}
\end{eqnarray}
where $m_{1}^{2}=1+2c_{1} +\hbar^2 l_4(l_4+2) + 2 \hbar^2 T(T+1) $, $ m_{2}^{2}=1+2c_{2} +\hbar^2 l_4(l_4+2) - 2 \hbar^2 T(T+1)$.
For the unirreps to be finite dimensional, we impose the following constraints on the structure function (\ref{pp1}),
\begin{eqnarray}
\Phi(p+1; u,E)=0,\quad \Phi(0;u,E)=0,\quad \Phi(x)>0,\quad \forall x>0,\label{pro2}
\end{eqnarray}
where $p$ is a positive integer. These constraints give $(p+1)$-dimensional unirreps and their solution gives the energy $E$ and constant $u$. The energy spectrum is
\begin{eqnarray}
E=-\frac{c_{0}^{2}}{2( p+1+\frac{m_{1}+m_{2}}{2})^{2}}.\label{en1}
\end{eqnarray}
The total number of degeneracies depends on $p+1$ only when the other quantum numbers would be fixed. These quantum numbers do not increase the total number of degeneracies.

\section{Separation of variables}
In this section we show that the Hamiltonian (\ref{KP1}) is multiseparatble and allows separation of variables in the hypersherical and parabolic coordinates. We also show that the dual system of (\ref{KP1}) is separable in the Euler and cylindrical coordinates.

\subsection{Hyperspherical coordinates}
We define the hyperspherical coordinates $r\in [0,\infty)$, $\theta\in[0,\pi]$, $\alpha\in[0,2\pi)$, $\beta\in[0,\pi]$ and $\gamma\in[0,4\pi)$ in the space $\mathbb{R}^5$ by
\begin{eqnarray}
&&x_0=r\cos\theta,\nonumber
\\&&
x_1+i x_2=r\sin\theta\cos\frac{\beta}{2}e^{i(\alpha+\gamma)/2},\nonumber
\\&&
x_3+i x_4=r\sin\theta\sin\frac{\beta}{2}e^{i(\alpha-\gamma)/2}.
\end{eqnarray}
In this coordinate system the differential elements of length, volume and the Laplace operator can be expressed \cite{Mad12} as
\begin{eqnarray}
&&dl^2=dr^2+r^2d\theta^2+\frac{r^2}{4}\sin^2\theta(d\alpha^2+d\beta^2+d\gamma^2+2\cos\theta d\alpha d\gamma),\nonumber\\&&
dV=\frac{r^4}{8}\sin^3\theta\sin\beta drd\theta d\alpha d\beta d\gamma,\nonumber\\&&
\Delta=\frac{1}{r^{4}}\frac{\partial}{\partial r} ( r^{4} \frac{\partial}{ \partial r}) + \frac{1}{r^{2}\sin^{3}\theta}\frac{\partial}{ \partial \theta}( \sin^{3}\theta\frac{\partial}{ \partial \theta})-\frac{4\hat{L}^2}{r^2\sin^2\theta},
\end{eqnarray}
with $\hat{L}^2=L_1^2+L_2^2+L_3^2 $ and
\begin{eqnarray}
&&L_1=i\left(\cos\alpha\cot\beta\frac{\partial}{\partial\alpha}+\sin\alpha\frac{\partial}{\partial\beta}-\frac{\cos\alpha}{\sin\beta}\frac{\partial}{\partial\gamma}\right),\nonumber\\&&
L_2=-i\left(\sin\alpha\cot\beta\frac{\partial}{\partial\alpha}-\cos\alpha\frac{\partial}{\partial\beta}-\frac{\sin\alpha}{\sin\beta}\frac{\partial}{\partial\gamma}\right),\nonumber\\&&
L_3=i\frac{\partial}{\partial\alpha}.
\end{eqnarray}
With the help of the identity\cite{Mad1}
\begin{eqnarray}
iA^a_j\frac{\partial}{\partial x_j}=\frac{2}{r(r+x_0)}L_a,
\end{eqnarray}
with
\begin{eqnarray}
&& L_{1}=\frac{i}{2}(D_{41}+D_{32}), \qquad L_{2}=\frac{i}{2}(D_{13}+D_{42}),\nonumber\\
&& L_{3}=\frac{i}{2}(D_{12}+D_{34}),\quad   D_{jk}=-x_{j}\frac{\partial}{\partial x_{k}}+x_{k} \frac{\partial}{\partial x_{j}},
\end{eqnarray}
the Schr\"{o}dinger equation $H\psi=E\psi$ of (\ref{KP1}) can be written as
\begin{eqnarray}
\left[\Delta_{r \theta}- \frac{\hat{L}^2+c_{2}}{\hbar^2 r^{2}\sin^{2}\frac{\theta}{2}}-\frac{\hat{J}^{2}+c_{1}}{\hbar^2 r^{2}\cos^{2}\frac{\theta}{2}} \right]\psi + \frac{2}{\hbar^{2}}(E+\frac{c_{0}}{r})\psi =0, \label{kp2}
\end{eqnarray}
where
\begin{eqnarray}
\Delta_{r \theta}=\frac{1}{r^{4}}\frac{\partial}{\partial r} ( r^{4} \frac{\partial}{ \partial r}) + \frac{1}{r^{2}\sin^{3}\theta}\frac{\partial}{ \partial \theta}( \sin^{3}\theta\frac{\partial}{ \partial \theta}),
\end{eqnarray}
$\hat{J}^2=J_aJ_a$ with $J_a=L_a+T_a$, $a=1, 2, 3$. The operators $L_a$ and $J_a$ satisfy the commutation relations
\begin{eqnarray}
[ L_{a},L_{b}]=i \epsilon_{abc} L_{c},\quad [ J_{a},J_{b}]=i \epsilon_{abc} J_{c}.
\end{eqnarray}
Equation (\ref{kp2}) is separable in the hyperspherical coordinates using the eigenfunctions of $\hat{L}^2$, $\hat{T}^2$ and $\hat{J}^2$ with the eigenvalues $\hbar^2 L(L+1)$, $\hbar^2 T(T+1)$ and $\hbar^2 J(J+1)$, respectively. This is seen as follows. We make the separation ansatz \cite{Mad1}
\begin{eqnarray}
\psi &=& \Phi(r,\theta) D^{JM}_{LTm't'}(\alpha,\beta,\gamma,\alpha_{T},\beta_{T},\gamma_{T}),
\end{eqnarray}
where
\begin{eqnarray}
 D^{JM}_{LTm't'}(\alpha,\beta,\gamma,\alpha_{T},\beta_{T},\gamma_{T})&= &\sqrt{\frac{(2L+1)(2T+1)}{4 \pi^{4}}} \sum_{M=m+t}
 C_{L,m;T,t}^{JM}\nonumber\\&&\times D_{mm'}^{L}(\alpha,\beta,\gamma) D_{tt'}^{T}(\alpha_{T},\beta_{T},\gamma_{T}),
\end{eqnarray}
and $C_{L,m;T,t}^{JM}$ are the Clebseh-Gordon coefficients that arise in angular momentum coupling and appear as the expansion coefficients of total angular momentum eigenstates in uncoupled tensor product basis \cite{Con1}; $D^j_{mm'}$ are the $su(2)$ Wigner D-functions of dimension $2j+1$ with $j=0, 1/2, 1, 3/2, 2,\dots $ and $m=-j, -j+1,\dots, j$ \cite{Wig1}.
Substituting the ansatz into (\ref{kp2}) leads to the differential equation for the the function $\Phi(r,\theta)$
\begin{eqnarray}
&&\left[\Delta_{r \theta}- \frac{L(L+1)+\frac{c_{2}}{\hbar^{2}}}{r^{2}\sin^{2}\frac{\theta}{2}}-\frac{J(J+1)+\frac{c_{1}}{\hbar^{2}}}{r^{2}\cos^{2}\frac{\theta}{2}}+ \frac{2}{\hbar^{2}}(E+\frac{c_{0}}{r}) \right]\Phi(r,\theta) =0.\nonumber\\&& \label{kpp1}
\end{eqnarray}
Setting the function $\Phi(r,\theta)=R(r)F(\theta)$, (\ref{kpp1}) is separated into the ordinary differential equations
\begin{eqnarray}
&&\left[\frac{d^2}{d\theta^2}+3\cot\theta\frac{d}{d\theta} - \frac{2\{L(L+1)+\frac{c_{2}}{\hbar^{2}}\}}{1-\cos\theta}
-\frac{2\{J(J+1)+\frac{c_{1}}{\hbar^{2}}\}}{1+\cos\theta}+\Lambda\right] F(\theta)=0, \nonumber\\
&&\label{kpp3}
\\&&
\left[\frac{d^2}{d r^2} +\frac{4}{r}\frac{d}{d r} + \frac{2}{\hbar^{2}}(E+\frac{c_{0}}{r})-\frac{\Lambda}{r^2}\right]R(r)=0,\label{kpp4}
\end{eqnarray}
where $\Lambda$ is the separation constant. Solutions of (\ref{kpp3}) and (\ref{kpp4}) in terms of the Jacobi and confluent hypergeometric polynomials \cite{And1} are as follows
\begin{eqnarray}
&&F(\theta)=F_{\lambda J L}(\theta;\delta_1,\delta_2)(1+\cos\theta)^{(\delta_{1}+J)/2}(1-\cos\theta)^{(\delta_{2}+L)/2}\nonumber\\&&\quad\quad\quad \times P_{\lambda -J-L}^{( \delta_{2}+L, \delta_{1}+J)}(\cos\theta),\nonumber
\\&&
R(r)=R_{n \lambda}(r:\delta_1,\delta_2)e^{-\frac{\kappa r}{2}} (\kappa r)^{\lambda + \frac{\delta_{1}+\delta_{2}}{2}} {}_1 F_1(-n, 4 +2 \lambda + \delta_{1}+\delta_{2};\kappa r),\nonumber\\&&
\end{eqnarray}
where
\begin{eqnarray}
&&\Lambda=(\lambda+\frac{\delta_1+\delta_2}{2})(\lambda+\frac{\delta_1+\delta_2}{2}+3),\quad \lambda\in \mathbb{N},\nonumber
\\&&\delta_{1}=-1+\sqrt{\frac{4c_{1}}{\hbar^{2}}+(2J+1)^{2}}-J,\nonumber
\\&&
\delta_{2}=-1+\sqrt{\frac{4c_{2}}{\hbar^{2}}+(2L+1)^{2}}-L,\nonumber
\\&&
-n= -\frac{2c_{0}}{\hbar^2 \kappa}+ \frac{\delta_{1}+\delta_{2}}{2} +\lambda+2, \quad E=\frac{-\kappa^2\hbar^2}{8}.
\end{eqnarray}
Hence the energy spectrum
\begin{eqnarray}
E=-\frac{c_{0}^{2}}{2 \hbar^{2}( n+\lambda+2+ \frac{\delta_{1}+\delta_{2}}{2})^{2}}.\label{en2}
\end{eqnarray}
This physical spectrum coincides with (\ref{en1}) obtained from algebraic derivation by the identification $p= n+\lambda+1$, $\delta_1=m_1$, $\delta_2=m_2$ and $\hbar=1$.

\subsection{Parabolic coordinates}
The parabolic coordinates are defined by
\begin{eqnarray}
&&x_0=\frac{1}{2}(\mu-\nu),\nonumber
\\&&
x_1+i x_2=\sqrt{\mu\nu}\cos\frac{\beta}{2}e^{i(\alpha+\gamma)/2},\nonumber
\\&&
x_3+i x_4=\sqrt{\mu\nu}\sin\frac{\beta}{2}e^{i(\alpha-\gamma)/2},
\end{eqnarray}
with $\mu,\nu\in[0,\infty)$. The differential elements of length, volume and Laplace operator in this coordinates can be expressed as
\begin{eqnarray}
&&dl^2=\frac{\mu+\nu}{4}\left(\frac{d\mu^2}{\mu}+\frac{du^2}{\nu}\right)+\frac{\mu\nu}{4}(d\alpha^2+d\beta^2+d\gamma^2+2\cos\theta d\alpha d\gamma),\nonumber\\&&
dV=\frac{\mu\nu}{32}(\mu+\nu)\sin\beta d\mu d\nu d\alpha d\beta d\gamma,\nonumber\\&&
\Delta=\frac{4}{\mu+\nu}\left[ \frac{1}{\mu}\frac{\partial}{\partial \mu} ( \mu^{2} \frac{\partial}{ \partial \mu})+\frac{1}{\nu}\frac{\partial}{\partial \nu} ( \nu^{2} \frac{\partial}{ \partial \nu})\right]-\frac{4}{\mu\nu}\hat{L}^2.
\end{eqnarray}
The Schr\"{o}dinger equation $H\psi=E\psi$ of the Hamiltonian (\ref{KP1}) in this coordinates becomes
\begin{eqnarray}
\left[ \Delta_{\mu \nu}-  \frac{ 4(\hat{J}^{2}+c_1)}{\hbar^2 \mu (\mu +\nu)} -  \frac{ 4(\hat{L}^2+c_2)}{\hbar^2\nu(\nu+\mu)}+ \frac{2}{\hbar^{2}}(E+\frac{c_{0}}{\mu+\nu})\right]\psi =0,\label{kp3}
\end{eqnarray}
where
\begin{eqnarray}
\Delta_{\mu \nu}=  \frac{4}{\mu+\nu}\left[ \frac{1}{\mu}\frac{\partial}{\partial \mu} ( \mu^{2} \frac{\partial}{ \partial \mu})+\frac{1}{\nu}\frac{\partial}{\partial \nu} ( \nu^{2} \frac{\partial}{ \partial \nu})\right].
\end{eqnarray}
Making the ansatz of the form \cite{Mad5}
\begin{eqnarray}
&&\psi = f_{1}(\mu) f_{2}(\nu) D_{LTm't'}^{JM}(\alpha,\beta,\gamma,\alpha_{T},\beta_{T},\gamma_{T}),
\end{eqnarray}
in (\ref{kp3}), the wave functions are separated and lead to the following ordinary differential equations
\begin{eqnarray}
&&\frac{1}{\mu} \frac{d}{d\mu}(\mu^2 \frac{ d f_{1}}{d\mu})+ \left[ \frac{E}{2\hbar^{2}} \mu  -  \frac{ J(J+1)+\frac{c_1}{\hbar^2}}{\mu}+\frac{c_0}{2\hbar^2}+\frac{\hbar}{2}\tilde{\Lambda} \right]f_{1}=0,\label{kp4}
\\&&
\frac{1}{\nu} \frac{d}{d\nu}( \nu^2 \frac{ d f_{2}}{d\nu})  +  \left[ \frac{E}{2\hbar^{2}} \nu  -  \frac{ L(L+1)+\frac{c_2}{\hbar^2}}{\nu}+\frac{c_0}{2\hbar^2}-\frac{\hbar}{2}\tilde{\Lambda}\right]f_{2}=0,\label{kp5}
\end{eqnarray}
where $\tilde{\Lambda}$ is the separation constant.
Solutions of (\ref{kp4}) and (\ref{kp5}) are given by the confluent hypergeometric polynomials \cite{And1},
\begin{eqnarray}
&&f_{1}=e^{-\frac{\kappa\mu}{2}}( \kappa \mu)^{\delta_{1}+J}F(-n_{1}, \delta_{1}+J +2 , \kappa \mu ),\nonumber
\\&&
f_{2}=e^{-\frac{\kappa\nu}{2}}( \kappa \nu)^{\delta_{2}+L}F(-n_{2}, \delta_{2}+L +2 , \kappa \nu ),
\end{eqnarray}
where
\begin{eqnarray}
&&\delta_{1}=-1+\sqrt{\frac{4c_{1}}{\hbar^{2}}+(2J+1)^{2}}-J,\nonumber
\\&&
\delta_{2}=-1+\sqrt{\frac{4c_{2}}{\hbar^{2}}+(2L+1)^{2}}-L,\nonumber
\\&&
-n_{1}=\frac{1}{2}(\delta_{1}+J)+1-\frac{\hbar}{2\kappa}\tilde{\Lambda} -\frac{c_{0}}{2\kappa \hbar^{2}},\nonumber
\\&&
-n_{2}=\frac{1}{2}(\delta_{2}+L)+1+\frac{\hbar}{2\kappa}\tilde{\Lambda} -\frac{c_{0}}{2\kappa \hbar^{2}},\quad E=\frac{-\hbar^2\kappa^2}{2}.
\end{eqnarray}
Set
\begin{eqnarray}
n_{1}+n_{2}=\frac{c_0}{\kappa\hbar^2}-\frac{1}{2}(\delta_{1}+\delta_{2}+J+L)-2.
\end{eqnarray}
The energy spectrum
\begin{eqnarray}
E=-\frac{c_{0}^{2}}{2\hbar^{2}\{n_1+n_2+\frac{1}{2}(\delta_1+\delta_2+J+L)+2\}^{2}}.
\end{eqnarray}
Making the identification $p= n_1+n_2+\frac{J+L}{2}+1$, $\delta_1=m_1$, $\delta_2=m_2$ and $\hbar=1$, the energy spectrum becomes (\ref{en1}).

\subsection{Euler spherical coordinates}
The 5D Kepler system with non-central terms and Yang-Coulomb monopole is dual to the 8D singular oscillator via Hurwitz transformation \cite{Ian1}.
The symmetry algebra structure and energy spectrum of the 8D singular oscillator has been studied in \cite{Ian1}. In this and next subsections
we show the separability this dual system in the Euler spherical and cylindrical coordinates, and compare our results for the spectrum with those obtained in \cite{Ian1}.

The Hamiltonian of the 8D singular oscillator reads
\begin{eqnarray}
&&H=-\frac{\hbar^{2}}{2}\sum_{i=0}^7 \frac{ \partial^{2}}{\partial u_i^{2}} + \frac{ \omega^{2}}{2}\sum_{i=0}^7 u_i^{2} +\frac{ \lambda_{1}}{u_{0}^{2}+u_{1}^{2}+u_{2}^{2}+u_{3}^{2}}+\frac{ \lambda_{2}}{u_{4}^{2}+u_{5}^{2}+u_{6}^{2}+u_{7}^{2}}.\nonumber\\&&\label{EP1}
\end{eqnarray}
In the Euler 8D spherical coordinates \cite{Kar1}
\begin{eqnarray}
&&u_{0}+iu_{1}=u \cos\frac{\theta}{2}\sin\frac{ \beta_{T}}{2} e^{-i\frac{(\alpha_{T}-\gamma_{T})}{2}},\nonumber
\\&&
u_{2}+iu_{3}=u \cos\frac{\theta}{2}\cos\frac{ \beta_{T}}{2} e^{i\frac{(\alpha_{T}+\gamma_{T})}{2}},\nonumber
\\&&
u_{4}+iu_{5}=u \sin\frac{\theta}{2}\sin \frac{ \beta_{K}}{2} e^{i\frac{(\alpha_{K}-\gamma_{K})}{2}},\nonumber
\\&&
u_{6}+iu_{7}=u\sin\frac{\theta}{2}\cos\frac{ \beta_K}{2} e^{-i\frac{(\alpha_{K}+\gamma_{K})}{2}},
\end{eqnarray}
where $0\leq u<\infty$, $0\leq \theta\leq\pi$, we have
\begin{eqnarray}
\sum_{i=0}^7 \frac{ \partial^{2}}{\partial u_i^{2}}&=&\frac{1}{u^{7}} \frac{\partial}{\partial u}( u^{7} \frac{\partial}{\partial u})+  \frac{4}{u^{2}\sin^3\theta}\frac{ \partial}{\partial \theta}( \sin^{3}\theta \frac{\partial }{ \partial \theta}) - \frac{4}{u^{2}\cos^{2}\frac{\theta}{2}}\hat{T}^2 - \frac{4}{u^{2}\sin^{2}\frac{\theta}{2}}\hat{K}^{2},\nonumber\\
\end{eqnarray}
with
\begin{eqnarray}
 \hat{T}^2 =-\left[\frac{\partial^{2}}{\partial \beta_{T}^{2}}+ \cot\beta_{T}\frac{\partial}{\partial \beta_{T}}+ \frac{1}{\sin^{2}\beta_{T}}\left( \frac{ \partial}{\partial \alpha_{T}^{2}} -2 \cos\beta_{T} \frac{\partial^{2}}{\partial \alpha_{T} \partial \gamma_{T}}+  \frac{\partial^{2}}{ \partial \gamma_{T}^{2}}\right)\right],
\end{eqnarray}
\begin{eqnarray}
 \hat{K}^{2} =-\left[\frac{\partial^{2}}{\partial \beta_{K}^{2}}+ \cot\beta_{K}\frac{\partial}{\partial \beta_{K}}+ \frac{1}{\sin^{2}\beta_{K}}\left( \frac{ \partial}{\partial \alpha_{K}^{2}} -2 \cos \beta_{K} \frac{\partial^{2}}{\partial \alpha_{K} \partial \gamma_{K}}+  \frac{\partial^{2}}{ \partial \gamma_{K}^{2}}\right)\right].
\end{eqnarray}
For the separation of the wavefunctions in the form
\begin{eqnarray}
\psi= R(u)F(\theta) D_{tt'}^{T}(\alpha_{T},\beta_{T},\gamma_{T})  D_{kk'}^{K}(\alpha_{K},\beta_{K},\gamma_{K}),
\end{eqnarray}
with
\begin{eqnarray}
\hat{T}^2D_{tt'}^{T}=T(T+1)D_{tt'}^{T},
\quad
\hat{K}^{2}D_{kk'}^{K}=K(K+1)D_{kk'}^{K},\label{WG1}
\end{eqnarray}
the Schr\"odinger equation $H\Psi=\epsilon \Psi$  leads to the ordinary differential equations
\begin{eqnarray}
&&\left[\frac{d^2}{d u^2}+ \frac{7}{u}\frac{d}{d u}- \frac{\Gamma}{u^{2}}+\frac{2 \epsilon}{\hbar^{2}}-\frac{\omega^{2}}{\hbar^{2}}u^{2}\right]R(u)=0,\label{rd1}
\\&&
\left[\frac{1}{\sin^{3}\theta} \frac{d}{d \theta}(\sin^{3}\theta\frac{d }{d \theta})- \frac{T(T+1)+\frac{\lambda_{1}}{2\hbar^{2}}}{  \cos^{2}\frac{\theta}{2}}- \frac{K(K+1)+\frac{\lambda_{2}}{2\hbar^{2}}}{  \sin^{2}\frac{\theta}{2}} + \frac{\Gamma}{4}\right] F(\theta) =0,\nonumber\\&&\label{ag1}
\end{eqnarray}
where $\Gamma$ is the separation of constant.
The solution of (\ref{ag1}) in terms of Jacobi polynomials \cite{And1} as follows
\begin{eqnarray}
&&F(\theta)=F_{\lambda T K}(\theta;\delta_1,\delta_2)(1+\cos\theta)^{(\delta_{1}+T)/2}(1-\cos\theta)^{(\delta_{2}+K)/2}\nonumber\\&&\quad\quad\quad \times P_{\lambda -T-K}^{( \delta_{2}+K, \delta_{1}+T)}(\cos\theta),
\end{eqnarray}
where
\begin{eqnarray}
&&\Gamma=4(\lambda+\frac{\delta_1+\delta_2}{2})(\lambda+\frac{\delta_1+\delta_2}{2}+3),\quad \lambda\in \mathbb{N},\nonumber
\\&&\delta_{1}=-1+\sqrt{\frac{2\lambda_{1}}{\hbar^{2}}+(2T+1)^{2}}-T,\nonumber
\\&&
\delta_{2}=-1+\sqrt{\frac{2\lambda_{2}}{\hbar^{2}}+(2K+1)^{2}}-K.
\end{eqnarray}
The solution of (\ref{rd1}) in terms of the confluent hypergeometric functions \cite{And1}
\begin{eqnarray}
R(u)=R_{n \lambda}(u;\delta_1,\delta_2)e^{-\frac{\kappa u^2}{2}} (\kappa u^2)^{\lambda + \frac{\delta_{1}+\delta_{2}}{2}} {}_1 F_1(-n, 4 +2 \lambda + \delta_{1}+\delta_{2};\kappa u^2),
\end{eqnarray}
where
\begin{eqnarray}
-n=\frac{\delta_1+\delta_2}{2}+\lambda+2-\frac{\epsilon}{2\kappa\hbar^2},\quad \omega^2=\kappa^2\hbar^2.
\end{eqnarray}
Hence
\begin{eqnarray}
\epsilon=2 \hbar^2 \kappa ( n +\frac{\delta_{1}+\delta_{2}}{2}+ \lambda +2 ).
\end{eqnarray}
Using the relations between the parameters of the generalized Yang-Coulomb monopole and the 8D singular oscillator,
\begin{eqnarray}
c_0=\frac{\epsilon}{4}, \quad E=\frac{-\omega^2}{8}, \quad 2c_i=\lambda_i,\quad i=1, 2,\label{duality}
\end{eqnarray}
we obtain
\begin{eqnarray}
E=-\frac{c_0^2}{2\hbar^2\left(n+\lambda + 2+\frac{\delta_1+\delta_2}{2}\right)^2},
\end{eqnarray}
which coincides with (\ref{en2}).

\subsection{Cylindrical coordinates}

Consider the 8D cylindrical coordinates
\begin{eqnarray}
&&u_{0}+iu_{1}= \rho_{1} \sin\frac{ \beta_{T}}{2} e^{-i\frac{(\alpha_{T}-\gamma_{T})}{2}},\nonumber
\\&&
u_{2}+iu_{3}= \rho_{1} \cos\frac{ \beta_{T}}{2}e^{i\frac{(\alpha_{T}+\gamma_{T})}{2}},\nonumber
\\&&
u_{4}+iu_{5}= \rho_{2}\sin\frac{ \beta_{K}}{2}e^{i\frac{(\alpha_{K}-\gamma_{K})}{2}},\nonumber
\\&&
u_{6}+iu_{7}= \rho_{2}\cos\frac{ \beta_{K}}{2}e^{-i\frac{(\alpha_{K}+\gamma_{K})}{2}},
\end{eqnarray}
where $0\leq \rho_1, \rho_2<\infty$. Then we have in this coordinate system,
\begin{eqnarray}
\sum_{i=0}^7 \frac{ \partial^{2}}{\partial u_i^{2}}= \frac{1}{\rho_{1}^{3}} \frac{\partial}{\partial \rho_{1}}( \rho_{1}^{3} \frac{\partial}{\partial \rho_{1}})  + \frac{1}{\rho_{2}^{3}} \frac{\partial}{\partial \rho_{2}}( \rho_{2}^{3} \frac{\partial}{\partial \rho_{2}})-\frac{4}{\rho_{1}^{2}}\hat{T}^2-\frac{4}{\rho_{2}^{2}}\hat{K}^{2},
\end{eqnarray}
\begin{eqnarray}
 \hat{T}^2
  =-\left[\frac{\partial^{2}}{\partial \beta_{T}^{2}}+ \cot\beta_{T} \frac{1}{\sin^{2}\beta_{T}}\left( \frac{ \partial}{\partial \alpha_{T}^{2}} -2 \cos\beta_{T} \frac{\partial^{2}}{\partial \alpha_{T} \partial \gamma_{T}}+  \frac{\partial^{2}}{ \partial \gamma_{T}^{2}}\right)\right],
\end{eqnarray}
\begin{eqnarray}
 \hat{K}^{2} =-\left[\frac{\partial^{2}}{\partial \beta_{K}^{2}}+ \cot\beta_{K} \frac{1}{\sin^{2}\beta_{K}}\left( \frac{ \partial}{\partial \alpha_{K}^{2}} -2 \cos \beta_{K} \frac{\partial^{2}}{\partial \alpha_{K} \partial \gamma_{K}}+  \frac{\partial^{2}}{ \partial \gamma_{K}^{2}}\right)\right].
\end{eqnarray}
Making the ansatz
\begin{eqnarray}
\psi = f_{1} f_{2} D_{tt'}^{T}(\alpha_{T},\beta_{T},\gamma_{T})  D_{kk'}^{K}(\alpha_{K},\beta_{K},\gamma_{K})
\end{eqnarray}
and change of variables $x_{i}= a^{2} \rho_{i}^{2}$, $a =\sqrt{\frac{\omega}{\hbar}}$, the Schr\"odinger equation $H\psi=\epsilon \psi$ is converted to
\begin{eqnarray}
&&x_{1} \frac{d^{2}f_{1}}{dx_{1}^{2}}+2 \frac{df_{1}}{dx_{1}}-\left[\frac{ T(T+1)+\frac{\lambda_1}{2\hbar^2}}{x_{1}}+\frac{x_{1}}{4}-\frac{\epsilon_{1}}{2\hbar \omega}\right]f_{1}=0,\label{hh1}
\\&&
x_{2} \frac{d^{2}f_{2}}{dx_{2}^{2}}+2 \frac{df_{2}}{dx_{2}}-\left[\frac{ K(K+1)+\frac{\lambda_2}{2\hbar^2}}{x_{2}}+\frac{x_{2}}{4}-\frac{\epsilon_{2}}{2\hbar \omega}\right]f_{2}=0.\label{hh2}
\end{eqnarray}
where $ \epsilon_{1}+\epsilon_{2}=\epsilon$. Setting $f_{i}= e^{-\frac{x_{i}}{2}} x_{i}^{\frac{\delta_i+z_i}{2}} W(x_{i})$, $ i=1, 2$, then (\ref{hh1}) and (\ref{hh2}) become
\begin{eqnarray}
x_{i} \frac{ d^{2}W(x_{i})}{dx_{i}^{2}} + ( \gamma_{i} -x_{i}) \frac{dW(x_{i})}{dx_{i}}- \alpha_{i} W(x_{i}) =0,\label{pk1}
\end{eqnarray}
where $ \alpha_i=\frac{\gamma_i}{2}-\frac{\epsilon_i}{2\hbar \omega}, \gamma_{i}=\delta_i+z_i+2$, $z_1=T$, $z_2=K$ and
\begin{eqnarray}
\delta_{i}=-1+\sqrt{\frac{2\lambda_{i}}{\hbar^{2}}+(2z_i+1)^{2}}-z_i,\quad i=1, 2.
\end{eqnarray}
Solutions of (\ref{pk1}) in terms of confluent hypergeometric polynomials \cite{And1} are given by
\begin{eqnarray}
 f_{n_{i},\gamma_{i}} = ( a \rho_{i})^{\gamma_{i}-2} e^{-\frac{a^{2} \rho_{i}^{2}}{2}} {}_1 F_1(-n_{i}, \gamma_{i}, a^{2} \rho_{i}^{2}),
\end{eqnarray}
where
\begin{eqnarray}
n_{i}=\frac{\epsilon_{i}}{2\hbar \omega}-\frac{1}{2}(\delta_i+z_i+2), \quad i=1,2.
\end{eqnarray}
We thus found the energy spectrum of the  8D singular oscillator
\begin{eqnarray}
\epsilon=2\hbar\omega\left(n_1+n_2+\frac{\delta_1+\delta_2}{2}+\frac{T+K}{2}+2 \right).
\end{eqnarray}
Using the relations (\ref{duality}) between the parameters of the generalized Yang-Coulomb monopole and the 8D singular oscillator, we obtain
\begin{eqnarray}
E=-\frac{c_0^2}{2\hbar^2\left(n_1+n_2+\frac{\delta_1+\delta_2}{2}+\frac{T+K}{2}+2\right)^2},
\end{eqnarray}
which coincides with (\ref{en1}) by making the identification $p= n_1+n_2+\frac{T+K}{2}+1$, $\delta_1=m_1$, $\delta_2=m_2$ and $\hbar=1$.

\section{Conclusion}
One of the results of this paper is the determination of the higher rank quadratic algebra structure $Q(3; L^{so(4)},T^{su(2)})\oplus so(4)$ in the 5D deformed Kepler system with non-central terms and Yang-Coulomb monopole interaction. The structure constants of the quadratic algebra $Q(3; L^{so(4)},T^{su(2)})$ contain Casimir operators of the $so(4)$ and $su(2)$ Lie algebras. The realization of this algebra in terms of deformed oscillator enable us to provide the finite dimensional unitary representations and the degeneracy of the energy spectrum of the model. We also connected these results with method of separation of variables, solution in terms of orthogonal polynomials and the dual under Hurwitz transformation which is a 8D singular oscillator.

It would be interesting to extend these results to systems in curved spaces, in particular to the 8D pseudospherical (Higgs) oscillator and its dual system involving monopole  \cite{Bel2}. Moreover, higher dimensional superintegrable models with monopole interactions are still relatively unexplored area. One of the open problems is to construct the non-central deformations of the known higher-dimensional models \cite{Men1,Kri1} and their symmetry algebras.

{\bf Acknowledgements:}
The research of FH was supported by International Postgraduate Research Scholarship and Australian Postgraduate Award. IM was supported by the Australian Research Council through a Discovery Early Career Researcher Award DE 130101067. YZZ was partially supported by the Australian Research Council, Discovery Project DP 140101492. He would like to thank the Institute of Theoretical Physics, Chinese Academy of Sciences, for hospitality and support.

\end{document}